\documentclass[11pt]{article} 
\usepackage{moriond,epsfig}

\def\Journal#1#2#3#4{{#1} {\bf #2}, #3 (#4)}

\def\PRD{{\em Phys. Rev.} D}
 
\def\MNRAS{{\em Mon. Not. R. Astron. Soc.}}
\def\apj{{\em Astrophys. J.}}

\def\be{\begin{equation}} 
\def\ee{\end{equation}}
\def\bea{\begin{eqnarray}} 
\def\eea{\end{eqnarray}}
\begin{document} 

\vspace*{4cm} 

\title{PROBING DARK ENERGY WITH CONSTELLATION-X}

\author{D.~A. RAPETTI$^{1}$, S.~W. ALLEN$^{1}$, CON-X FACILITY SCIENCE TEAM}

\address{$^{1}$Kavli Institute for Particle Astrophysics and
Cosmology, Stanford University, 382 Via Pueblo Mall, \\ Stanford
94305-4060, USA.}

\maketitle\abstracts{Constellation-X (Con-X) will carry out two
powerful and independent sets of tests of dark energy based on X-ray
observations of galaxy clusters, providing comparable accuracy to
other leading dark energy probes. The first group of tests will
measure the absolute distances to clusters, primarily using
measurements of the X-ray gas mass fraction in the largest,
dynamically relaxed clusters, but with additional constraining power
provided by follow-up observations of the Sunyaev-Zel'dovich (SZ)
effect. As with supernovae studies, such data determine the
transformation between redshift and true distance, $d(z)$, allowing
cosmic acceleration to be measured directly. The second, independent
group of tests will use the exquisite spectroscopic capabilities of
Con-X to determine scaling relations between X-ray observables and
mass.  Together with forthcoming X-ray and SZ cluster surveys, these data will help to
constrain the growth of structure, which is also a strong function of
cosmological parameters.}

\section{Introduction}

The late-time accelerated expansion of the Universe is now a well
measured fact \cite{Riess:04,Allen:04,Spergel:06}. However, the
underlying cause of this cosmic acceleration remains unknown. Assuming
general relativity, a new energy component of the Universe, so-called
dark energy, is required. In order to pin down its nature, a number of
powerful future experiments are planned.

Con-X~\footnote{\sf http://constellation.gsfc.nasa.gov/} data will
constrain dark energy with comparable accuracy and in a beautifully
complementary manner to the best other techniques available circa
2018. Using a modest $\sim10-15\%$ (10-15Ms) investment of the
available observing time over the first 5 years of the Con-X mission,
we will be able to measure the X-ray gas mass fraction (or predict the
Compton $y$-parameter) to $5\%$ or $3.5\%$ accuracy, corresponding to
$3.3\%$ or $2.3\%$ in distance, for 500 or 250 clusters, respectively,
with a median redshift $z\sim1$.  When combined with CMB data or
suitable priors, the predicted dark energy constraints from Con-X
X-ray data are comparable to those projected by
e.g. future supernovae, weak lensing and baryon oscillation
experiments. Only by combining such independent and complementary
methods can a rigorous and precise understanding of the nature of dark
energy be achieved.

\section{Gas mass fraction, $f_{\rm gas}$, in X-ray galaxy clusters}

\begin{figure}
\begin{center}
\includegraphics[width=2.5in]{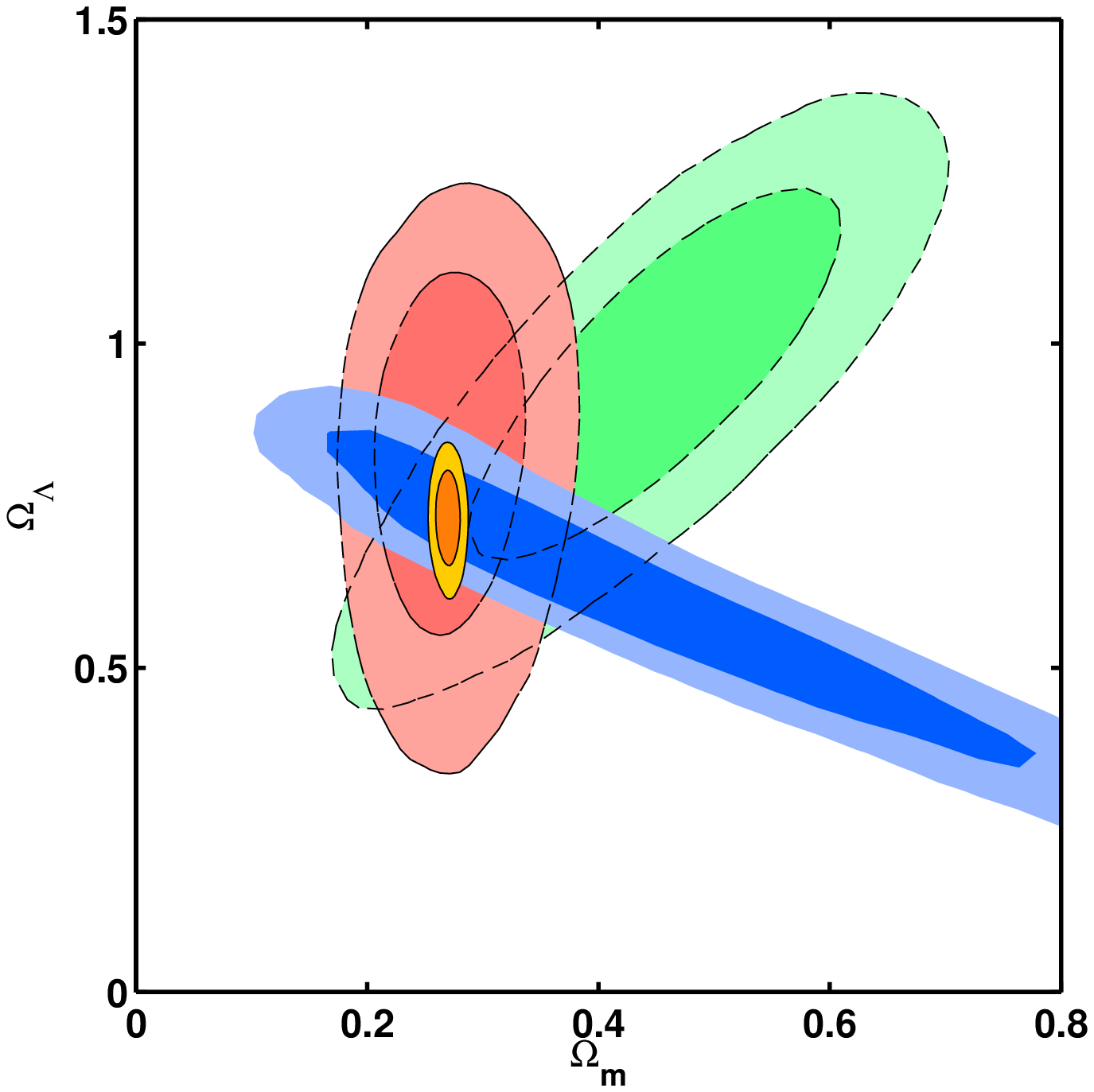}
\hspace{0.7cm}\includegraphics[width=2.45in]{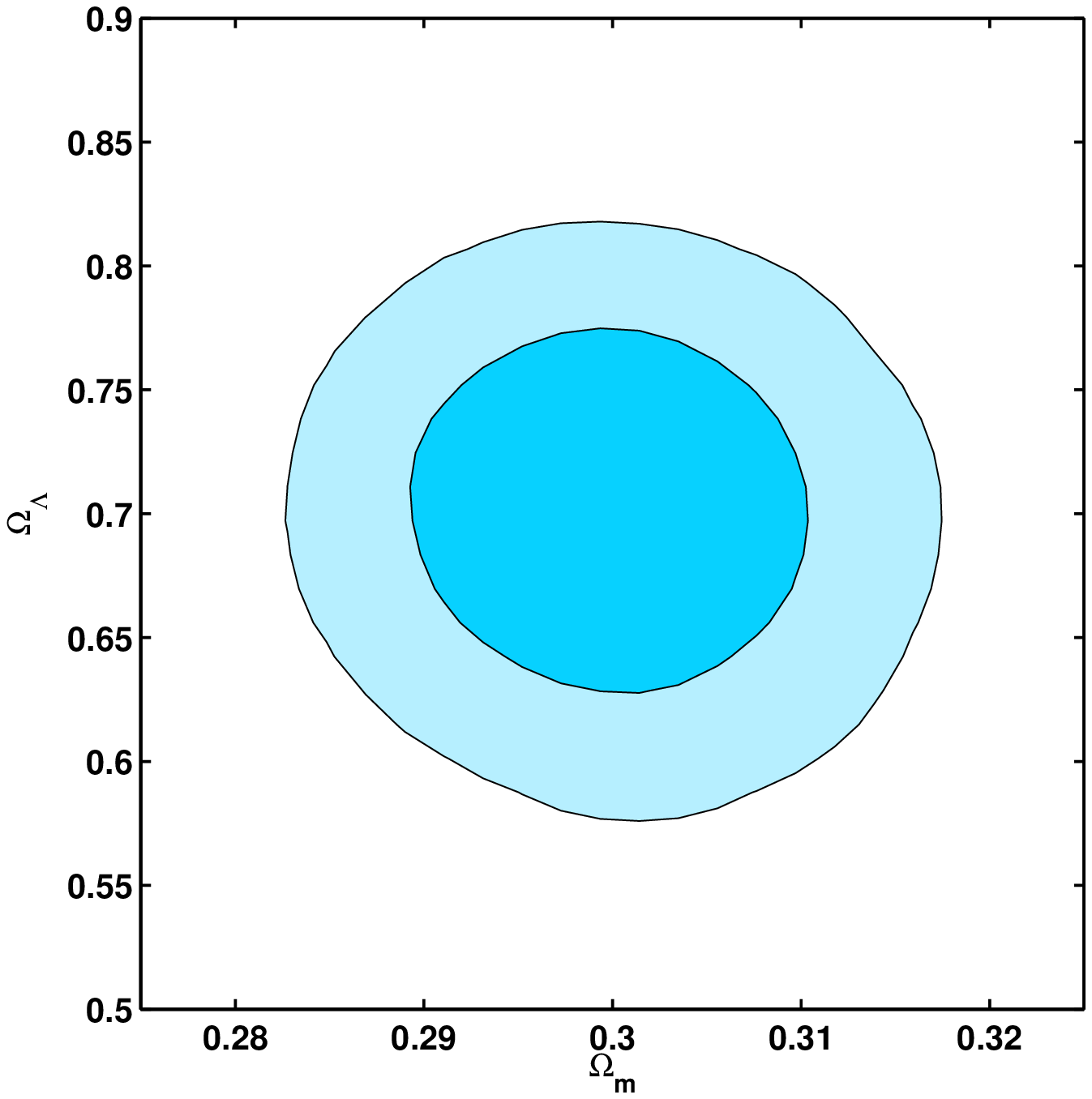}
\end{center}
\caption{(left panel) The joint 1 and 2 $\sigma$ contours on
  $\Omega_{\rm m}$ and $\Omega_{\rm \Lambda}$ from the current Chandra
  $f_{\rm gas}(z)$ data (red; Allen et. al 2006, in preparation). Also
  shown are the constraints from SNIa (green; Riess et al. 2004) and
  CMB studies (blue; 1st year WMAP+CBI+ACBAR). The same contours from the
  analysis of the simulated Con-X $f_{\rm gas}$ data set of
  Fig.~\ref{fig:simuldata} are shown in orange. These constraints are
  also shown alone in the right panel of this figure.}
\label{fig:comparison}
\end{figure}

The matter content of the largest, dynamically relaxed galaxy clusters
is expected to provide an almost fair sample of the matter content of
the Universe, $\Omega_{\rm m}$ \cite{White:91,White:93,Eke:98}. The
ratio of baryonic to total mass in such clusters should closely match
the ratio of the cosmological parameters $\Omega_{\rm b}/\Omega_{\rm
m}$ (where $\Omega_{\rm b}$ is the mean baryonic matter density of the
Universe in units of the critical density). Measurements of the ratio
of X-ray gas mass-to-total mass ratio in clusters as a function of
redshift,  $f_{\rm gas}(z)$, 
can also be used to measure cosmic acceleration directly \cite{Riess:04,Allen:04,Rapetti:06}.

Current Chandra X-ray Observatory $f_{\rm gas}(z)$ data for 41 hot
($kT > 5$keV), X-ray luminous ($L_{\rm X} > 10^{45}h^{-2}_{70}$ erg/s), 
dynamically relaxed clusters spanning the redshift range
$0.06 < z < 1.07$, provide a tight constraint on $\Omega_{\rm
m}=0.27\pm0.04$ and a $>99.99\%$ significant detection of the effects
of dark energy (cosmic acceleration) on the distances to the clusters 
\cite{Allen:06} (Figure~\ref{fig:comparison}:
red contours).  These selection criteria, especially the restriction to
the most relaxed systems, are essential to minimize systematic scatter
in the experiment \cite{Allen:04}. (Using these selection criteria, 
systematic scatter is undetected in the present Chandra data, for which
the unweighted rms $f_{\rm gas}$ measurement errors are 
$\sim 10\%$, corresponding to $\sim 7\%$ in distance.)
Rapetti {\it et al} (2005a,b)
discuss the complementary nature of $f_{\rm gas}$, CMB and type Ia
supernovae experiments and the impressive combined degeneracy-breaking
power of the data for dark energy studies.  Measurements of the growth
of structure through X-ray and SZ cluster surveys, weak lensing
measurements and baryon oscillation experiments also offer powerful
avenues of investigation.

The observed $f_{\rm gas}(z)$ values for a chosen reference (eg $\Lambda$CDM) cosmology
can be fitted with a model that accounts for the expected apparent
variation in $f_{\rm gas}(z)$ as the true, underlying cosmology is
varied

\begin{equation}
f_{\rm gas}^{\rm \Lambda CDM}(z) = \frac{ b\, \Omega_{\rm b}} {\left(1+0.19
\sqrt{h}\right) \Omega_{\rm m}} \left[ \frac{d_{\rm
A}^{\rm \Lambda CDM}(z)}{d_{\rm A}^{\rm de}(z)} \right]^{1.5}\;,
\label{eq:fgas}
\end{equation}

\noindent where $d_{\rm A}^{\rm de}(z)$ and $d_{\rm A}^{\rm \Lambda
  CDM}(z)$ are the angular diameter distances ($d_{\rm A}=d_{\rm
  L}/(1+z)^2$) to the clusters for a given dark energy $(de)$ model
and the reference $\Lambda$CDM cosmology, respectively.
$H_{\rm 0} = 100\,h\,{\rm km}\,{\rm sec}^{-1}\, {\rm Mpc}^{-1}$ is the
present Hubble parameter and $b$ is the `bias' factor by which the
baryon fraction is depleted with respect to the universal mean (a
small amount of gas is expelled by shocks when the cluster forms). 
In present work, the optically luminous baryonic mass in clusters 
is assumed to scale as 
$0.19h^{0.5}$ times the X-ray gas mass  \cite{Fuku:98,Allen:02} . 

The expected value of $b$ at a given mean enclosed mass overdensity
in a cluster can be calibrated by numerical simulations (being 
one of the more straightforward quantities for such simulations to predict). 
Current simulations suggest that a Gaussian prior on
$b=0.824\pm0.089$ may be appropriate for relaxed clusters in the 
very high mass/luminosity/temperature 
range studied \cite{Allen:04}. This includes a combined $10\%$ allowance 
for systematic uncertainties 
in the overall normalization from various contributing sources. 
Systematic evolution in the baryonic mass fraction 
with redshift is a potential source of systematic uncertainty.
Such evolution must be understood at the few $\%$ level if 
the full power of future X-ray data is
to be extracted. Efforts to provide 
improved numerical simulations for large samples 
of the largest, dynamically relaxed clusters, are underway.
Efforts to obtain precise, direct measurements
of the optically luminous baryonic mass in clusters and its 
evolution using deep ground-based imaging are also ongoing.

\begin{figure}
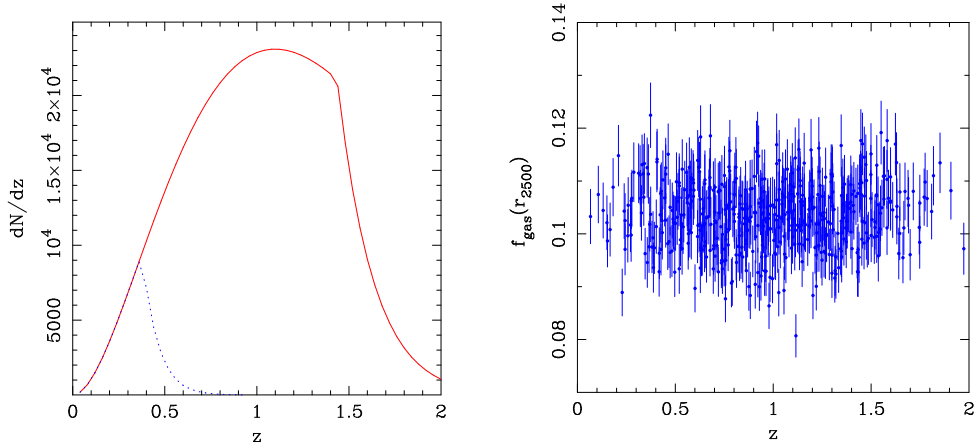

\begin{center}
\includegraphics[angle=270,width=2.3in]{dndz.ps}
\hspace{0.7cm}\includegraphics[angle=270,width=2.44in]{fgas_z_conx_2.ps}
\end{center}
\caption{(left panel) The predicted number density of clusters with
(bolometric) X-ray luminosities $L_{\rm X} > 10^{45}h_{70}^{-2}$erg/s for
a cluster survey flux limit of $5\times10^{-14}$erg\,cm$^{-2}$s$^{-1}$
in the $0.1-2.4$keV band (red, solid curve). The blue, dotted curve
shows the results for the same luminosity limit but a $0.1-2.4$keV 
flux limit of
$10^{-12}$erg\,cm$^{-2}$s$^{-1}$, appropriate for the ROSAT All-Sky
Survey. For illustration purposes, zero scatter
in the mass-luminosity relation and negligible flux errors are assumed. The
predictions are normalized to observations
at lower redshifts (Ebeling et al. 1998,
2001). (right panel) Projected $f_{\rm gas}(z)$ values for a possible
Con-X survey of 500 clusters with individual $f_{\rm gas}$ measurement
uncertainties of $5\%$. A systematic scatter of $4\%$ due to
cluster-cluster variations is included.}
\label{fig:simuldata}
\end{figure}

\section{Dark energy constraints from Con-X}

\begin{figure}
\begin{center}
\includegraphics[width=2.4in]{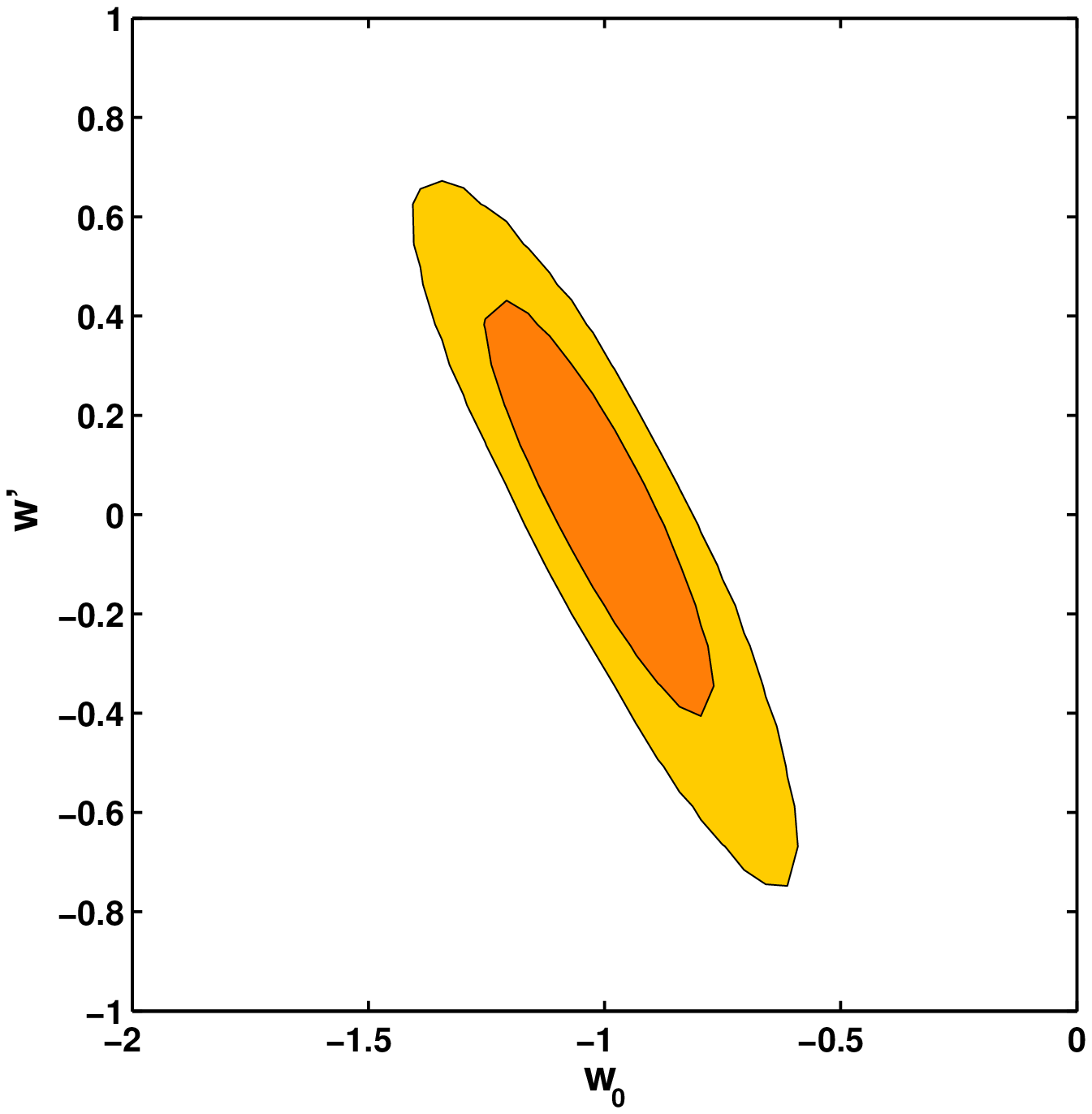}
\hspace{0.7cm}\includegraphics[width=2.5in]{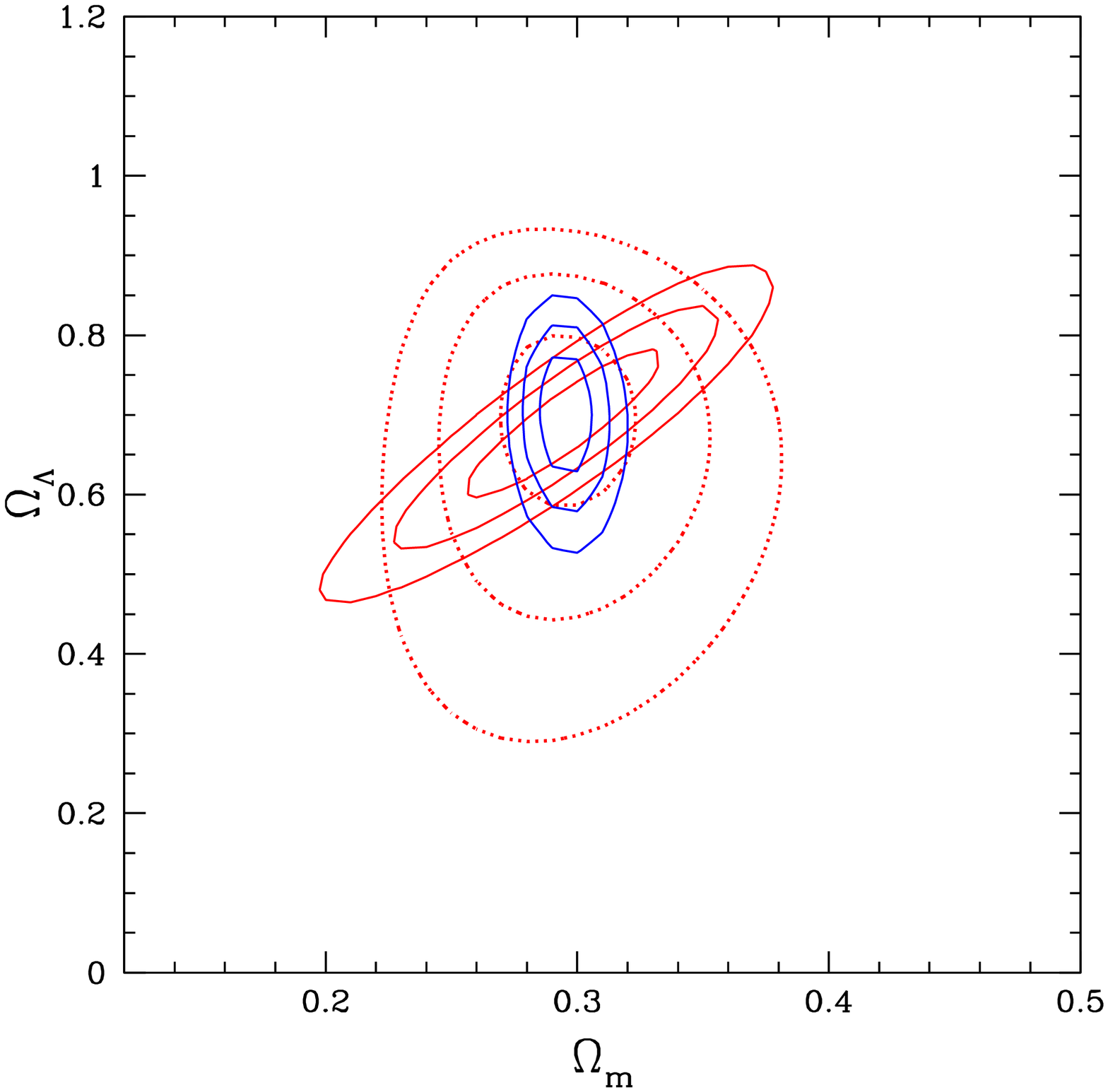}
\end{center}
\caption{(left panel) Results from the Markov Chain Monte Carlo (MCMC)
analysis of simulated Con-X $f_{\rm gas}$+CMB data. The CMB data are
WMAP TT data only, appropriate for 8 years of that mission (Upadhye et
al 2005). The 68.3 and 95.4 per cent uncertainties in the $(w_{\rm 0},
w')$ plane for a $w(a) = w_{\rm 0} + 2w' (1 - a)$ model are shown. 
A $2\%$ Gaussian prior on $b$ is assumed. No
priors on $\Omega_{\rm b} h^{2}$, $h$ or $\Omega_{\rm k}$ are required.  
Similar constraints are achievable from the $f_{\rm
gas}$ data alone, given $2\%$ priors on $\Omega_{\rm b} h^{2}$, $h$
and assuming flatness. Doubling the uncertainty in $b$ and/or allowing 
for systematic
evolution with redshift at the $\leq 4\%$ level does not worsen the 
plotted constraints significantly (see text for details).
(right panel) The 1, 2
and 3 $\sigma$ constraints for the Con-X X-ray+SZ experiment for the
500 cluster sample of Fig.~\ref{fig:simuldata} with $5\%$ statistical
errors in the predicted Compton $y$-values. The blue curve shows the
result from the $f_{\rm gas}$ experiment, as in
Fig.~\ref{fig:comparison} (orange). The dotted red curve shows the
results assuming a combined overall $2\%$ systematic uncertainty in
the normalization of the X-ray and SZ $y$-values and the solid red
curve when the systematic uncertainty in this and $h$ (combined) is
reduced to $0.1\%$.}
\label{fig:w}
\end{figure}

A possible Con-X $f_{\rm gas}$ study could involve a modest 10-15Ms
investment of observing time (approximately 10 per cent of the
available observing time over the first 5 years of the mission). Using
existing (at that time) X-ray and SZ cluster catalogs, one could
initially carry out short (5-10ks) Con-X snapshot exposures and/or
high resolution SZ observations of several thousand suitably
luminous/massive systems. Based on these initial results, one could
identify the 250-500 largest, relaxed clusters and re-observe these
for a further 10Ms total, with typical Con-X exposure times of 20 or
40 ks, leading to individual statistical error bars in $f_{\rm gas}$
measurements of 5 or 3.5\%.  (This gives a precision in the individual
distance measurements to the clusters of 3.3 or 2.3\%.)
Figure~\ref{fig:simuldata} shows the projected $f_{\rm gas}$
sample. We include a (possibly conservative) $4\%$ systematic scatter
in $f_{\rm gas}$ from cluster to cluster. (Current Chandra data
suggest the weighted mean systematic scatter to be $<5\%$
\cite{Allen:06}.)  We analyze this simulated Con-X $f_{\rm gas}$ data
set either alone (imposing $2\%$ Gaussian $1\sigma$ width priors on
$\Omega_{\rm b}h^{2}$ and $h$; Fig.~\ref{fig:comparison}, orange
contours) or combined with a simulated CMB data set (in this case 8
years of WMAP data) as shown in Figure~\ref{fig:w}.  We intitially use
a $2\%$ Gaussian prior on $b$ and assume, in the first instance,
negligible systematic evolution in the cluster baryonic mass
content with redshift.  We employ a full MCMC analysis in order to
properly explore the degeneracy breaking power.  For the evolving dark
energy case of Fig.~\ref{fig:w} we obtain comparable results to those
projected by supernovae \cite{Linder:04} and baryon oscillation
experiments \cite{Linder:05}. Allowing for unknown linear evolution in
the baryonic mass content of clusters at the 2\% level over the
redshift range $0<z<2$ has little effect on the results, increasing only
the uncertainity on $w'$ by $\sim 15\%$. Doubling the uncertainty in the evolution to
$4\%$ still has a negligible effect on $w_0$ but increases the
uncertainty on $w'$ by a further $\sim 25\%$. Relaxing the prior on
$b$ (the normalization) by a factor 2 ($4\%$ Gaussian
$1\sigma$ width) has little effect on $w_0$ but weakens the
constraints on $w'$ and $\Omega_{\rm m}$ by $\sim 15$ and $\sim 30\%$,
respectively.

The combination of X-ray and Sunyaev-Zel'dovich (SZ) data provides a
second, independent method by which to measure absolute distances to
clusters. The observed SZ flux can be expressed in terms of the
Compton $y$-parameter. This same Compton $y$-parameter can be
predicted from the X-ray data ($y_{\rm mod} \propto \int{n_e T dl}$),
with the predicted value depending on the assumed cosmology. For the
correct cosmology, the observed and predicted $y$-parameters should
agree \cite{Molnar:02,Molnar:04,Schmidt:04}. The right panel of
Figure~\ref{fig:w} shows the projected constraints for a standard
$\Lambda$CDM cosmology using the same 500 cluster Con-X sample and 
two illustrative SZ scenarios.

Clusters of galaxies are sensitive probes of the rate at which cosmic
structure evolves. Their number density at a fixed mass is
exponentially sensitive to the amplitude of linear matter density
perturbations. Measurements of the cluster mass function at different
redshifts constrains the perturbation growth parameter, which
is a second crucial dark energy observable.
Statistically, detailed studies of a sample of 1000 clusters can
constrain the growth factor to better than $0.5\%$, leading to
constraints on $w_{\rm 0}$ to $\pm 0.06-0.08$
\cite{Majumdar:04,Vikhlinin:05,Haiman:05}. The primary contribution of
Con-X to this work will be the precise calibration of cluster mass
measurements.

\section{Closing comments}

In probing cosmology via direct distance measurements, Con-X will
offer a powerful complement to SNIa studies, giving comparable
precision, but using different techniques and assumptions. Features to
note include $(i)$ that large relaxed galaxy clusters can, in
principle, be well modelled by simulations, although improvements in
this and other areas will be required to make full use of Con-X
data; $(ii)$ clusters are (in human terms) steady sources and can be
revisited to build up signal-to-noise on individual targets and
explore systematic issues; $(iii)$ the $f_{\rm gas}$
technique includes inbuilt complementary constraints on $\Omega_{\rm
m}$ from both the normalization and shape of the $f_{\rm gas}$ curve;
$(iv)$ the combination of $f_{\rm gas}$+CMB data breaks additional,
key cosmological parameter degeneracies in a remarkably effective
manner \cite{Rapetti:05a,Rapetti:06}; $(v)$ The systematic scatter in
the $f_{\rm gas}(z)$ data is small (undetected in current Chandra
data) once an appropriate restriction to large, relaxed clusters is
employed; $(vi)$ direct checks on assumptions like the form of
$b(z)$ are possible via combination with SZ data; this combination
also provides important extra cosmological information; (vii) the $f_{\rm gas}$ and X-ray+SZ experiments 
complement `cluster counting' experiments in that they do not
require complete samples (one can simply `cherry pick' the easiest
clusters to work with) and do not rely on calibration relations
to link observables to mass.

\section*{Acknowledgments}

This work was supported in part by the U.S. Department of Energy under
contract number DE-AC02-76SF00515. 

\section*{Note in press}

A preprint by E. Linder (arXiv:astro-ph/0606602) appears to significantly 
underestimate the precision of individual Con-X $f_{\rm gas}$ measurements 
and their degeneracy breaking power, as measured in full MCMC simulations. 
Our results will be discussed in more detail in a future paper.

\end{document}